# VERTICAL DUST PARTICLE CHAINS - MASS AND CHARGE MEASUREMENTS

Jorge Carmona-Reyes, Jimmy Schmoke, Mike Cook, Jie Kong and Truell W. Hyde[*]

CASPER, One Bear Place 97310
Baylor University, Waco, TX 76798-7310 USA

*Abstract*

Dusty plasmas have generated a large amount of interest since the discovery of ordered structure (crystal) formation in experimentally generated complex plasmas in 1994. Dust particles within the plasma sheath generated in these complex plasmas can form vertical chains due to the streaming ion wakefield. For the simplest of these configurations (a two particle chain), the particle closest to the lower electrode will generally remain in the shadow of the particle farthest from the lower electrode. These results in the two particles feeling differing ion drag forces: the top particle is acted on by the ion drag force directed from the plasma to the lower electrode, while the bottom particle is acted upon by the resulting 'wakefield' produced by the interaction of the upper particle with the ion drag force. This dynamic situation currently provides the best known experimental environment for examining the physics behind the ion drag force and its interaction with the plasma sheath. An experimental method for investigating the interaction between pair-particle chains based on modulating the bias on the lower electrode employing a DC bias modulation technique will be presented.

## I. INTRODUCTION

Complex plasmas are now considered an established area of research as illustrated by the increasing number of publications in the field. However, there is still much to do before a proper understanding of the physics behind the behavior of a dust grain or dust grain ensemble within a complex plasma environment is properly established. Such an understanding is of interest to both the plasma processing community due to the dust created instabilities produced during sputtering, etching and plasma deposition and the scientific community due to plasma-dust interactions observed in planetary ring systems, interstellar clouds and cometary environments [1].

Several different procedures have been proposed to measure the charge on a dust grain; common examples include the vertical resonance method, the dust particle equilibrium method and dust lattice waves [2]. This paper proposes a new mechanism employing the vertical alignment of a chain of dust particles. Employing an rf-discharge plasma and melamine formaldehyde particles of diameter 8.89 μm ± .09 μm, a chain of vertically aligned dust particles can be formed as shown in Fig. 1.

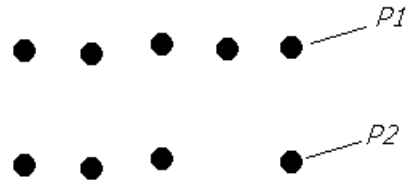

**Figure 1.** Normative experimental side view of an aligned particle chain system where the bottom particle (P2) is influenced by the wakefield created by the top particle (P1).

For a two-particle chain, once the dust particles reach equilibrium due to the balance of the gravitational and electrostatic forces, they align at the point where the lower particle feels the streaming of the ion wakefield and the top particle is bound by the negative ion drag force [3]. This experimentally observed phenomenon [4] provides a mechanism for calculating both the mass and the charge of the particle(s) and might ultimately allow a direct determination of the magnitude of the streaming ion wakefield through the proportional relationship existing between the resonant frequency and the charge to mass ratio of the two-particle system[5][6].

The behavior of such a vertical chain of particles can be described by the standard damped harmonic oscillation equations driven by a forcing term consisting of an applied DC modulation [7],

$$\ddot{x} + \beta \dot{x} + \omega_0^2 x = A \cos \omega t \quad (1)$$

where

$$\omega_0^2 = Q_D \frac{n_i e}{\varepsilon_0 m_D} \cdot \quad (2)$$

In the above, $Q_d$ is the charge on the particle, $m_d$ its mass and $\omega_0^2$ the square of the resonant frequency. When employing a center of mass system, $\omega_o$ will be related to the particles' mass and charge while the phase difference between the resonant frequencies of the individual particles will be related to the streaming ion wakefield assuming Bohm's criteria is met. This provides an experimental measurement mechanism, albeit with several new complications. The next section explains the theoretical foundation behind this experiment.

---

[*] email: Truell_Hyde@baylor.edu

## II. APPARENT MASS DIFFERENCE OF TWO PARTICLES

The primary forces experienced by a particle immersed within a complex plasma are given in Equation (3). Only non-negligible forces important to the self-assembly process are included. The potential inside the sheath is assumed to be parabolic to first order (i.e., $\phi \propto y^2$) and thus can be included as shown below.

$$m\ddot{y} = \Sigma F_i = -mg + QE - k\dot{y} + f_{ion} + f_{yukawa} + f_{wake} \quad (3)$$

$$m\ddot{y} + k\dot{y} + Q\frac{\partial \phi}{\partial y} = -mg + f_{yukawa} + f_{wake} \quad (4)$$

$$\ddot{y} + \beta\dot{y} + \omega_0^2 y = \frac{f_{Yukawa}}{m} + \frac{f_{wake}}{m} \quad (5)$$

$$\ddot{y} + \beta\dot{y} + \omega_0^2 y = \Sigma F_{rest} \quad (6)$$

Recasting the forces on the RHS of equation 6 provides the governing equations of motion for the particles as shown in equations 7a and 7b

$$\ddot{y}_1 + \beta_1\dot{y}_1 + \omega_{01}^2 y_1 = f_1(|y_1 - y_2|) \quad (7a)$$

$$\ddot{y}_2 + \beta_2\dot{y}_2 + \omega_{02}^2 y_2 = f_2(|y_1 - y_2|) \quad (7b)$$

Assuming a Maxwellian distribution of the neutral gas inside the chamber, the drag force will be the same for both particles; thus we can assume $\beta_1 = \beta_2 = \beta$ if in fact the difference in surface area between particles is minimal assuming mono-disperse particles. The resonant frequencies of particle 1 and 2 are $\omega_{01,2}$ respectively and the resultant force includes both the Yukawa and wake field forces. As shown, both these forces are a function of the distance between the two particles, i.e., $|y_1 - y_2|$.

Shifting equations (7a and 7b) from a particle frame of reference to a center of mass frame yields equations 8a and 8b,

$$(\ddot{Y}_{CM} + \beta\dot{Y}_{CM} + \omega_{01}^2 Y_{CM}) + \left(\frac{\gamma}{1+\gamma}\right)(\ddot{Y}_R + \beta\dot{Y}_R + \omega_{01}^2 Y_R) = f_1(Y_R) \quad (8a)$$

$$(\ddot{Y}_{CM} + \beta\dot{Y}_{CM} + \omega_{02}^2 Y_{CM}) + \left(\frac{1}{1+\gamma}\right)(\ddot{Y}_R + \beta\dot{Y}_R + \omega_{02}^2 Y_R) = f_2(Y_R) \quad (8b)$$

Once separated as shown below, the resonant frequency of the center of mass can be seen to be related to both the resonant frequency of the individual particles and their masses as given in equation 11b.

$$\ddot{Y}_{CM} + \beta\dot{Y}_{CM} + \omega_{01}^2 Y_{CM} = 0 \quad (9a)$$

$$\ddot{Y}_{CM} + \beta\dot{Y}_{CM} + \omega_{02}^2 Y_{CM} = 0 \quad (9b)$$

and

$$\left(\frac{\gamma}{1+\gamma}\right)(\ddot{Y}_R + \beta\dot{Y}_R + \omega_{01}^2 Y_R) = f_1(Y_R) \quad (10a)$$

$$-\left(\frac{1}{1+\gamma}\right)(\ddot{Y}_R + \beta\dot{Y}_R + \omega_{02}^2 Y_R) = f_2(Y_R) \quad (10b)$$

$$\ddot{Y}_R + \beta\dot{Y}_R + \Omega_0^2 Y_R = F(Y_R) \quad (11a)$$

$$\Omega_0^2 = \frac{\gamma\omega_{01}^2 + \omega_{02}^2}{1+\gamma} \quad (11b)$$

Solving for $\gamma$ in Equation (11b), yields

$$\gamma = \frac{\omega_{02}^2 - \Omega_0^2}{\Omega_0^2 - \omega_{01}^2} \quad (12)$$

Thus, any apparent mass difference can be determined once the individual resonance frequencies $\omega_{01,2}$ and the center of mass resonance frequency $\Omega_0$ are obtained experimentally.

## III. EXPERIMENTAL SETUP

In order to calculate the ratio of the two masses ($\gamma$), experimentally determined values for $\omega_{01,2}$, $\Omega_0$ and $\beta$ must be obtained. In this experiment, MF Micro spheres of 8.89 μm ± .09 μm in diameter [http://www.microparticles.de/cgi-bin/produktengl.cgi] were introduced into CASPER's GEC reference cell [8]. An argon plasma was ignited using an rf frequency generator at 13.56 MHz, producing a natural negative bias. Pressure within the cell was held at 13.33 Pa, a flow rate of 20 sccm's was established and 10 Watts of power delivered to the system. A plate with an integrated one-inch cutout was placed on the lower electrode to create the necessary boundary conditions for forming the coulomb crystal. In order to maintain consistency, reflected power was held constant at 10% of the forward power delivered to the system. As shown in Fig. 2, the vertical layers of the cluster can be clearly seen where these particles are illuminated employing a 660 nm, 80mW diode laser.

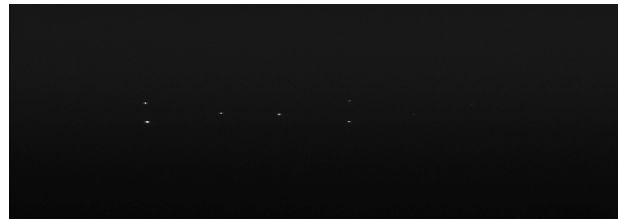

**Figure 2.** View of aligned particles.

Once the chain stabilized, an externally modulated DC bias was provided to the cell's lower electrode via a KEPCO operational power supply [9] as shown in Figure 3. The filter shown only allows signals below 500 Hz to pass, optimizing the DC modulation technique used in these experiments. A modulated, external DC bias comprised of an arbitrary waveform ranging from 5 Hz to 35 Hz as shown in Fig. 4 was then fed to a 33120A Agilent signal generator via a DTE (Data Terminal Equipment cable). A diagram of the entire system is given in Fig. 5.

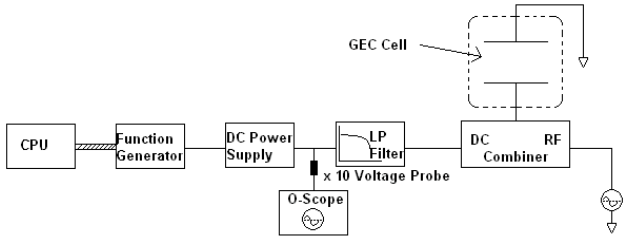

**Figure 3.** Representative modulated DC waveform.

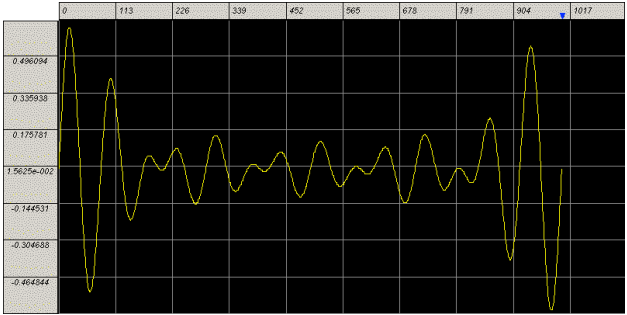

**Figure 4.** Sample of a representative signal fed to the DC power supply.

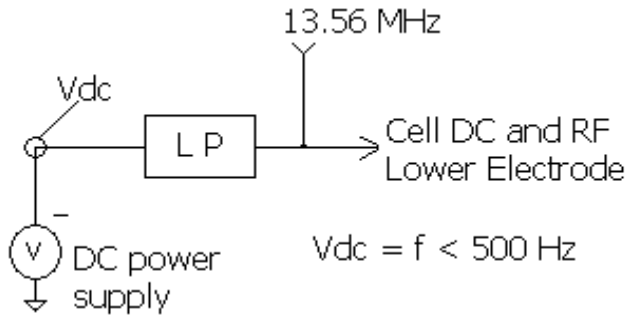

**Figure 5.** Diagram illustrating the manner in which an external DC power supply is integrated with the cell.

The signal used was created employing the Agilent IntuiLink Waveform editor Version 1.4.5. Using a randomly generated wave function from the signal generator, the created signal was fed into the waveform editor, translated and then converted into a signal the function generator could deliver. The final waveform was composed via MatLAB using its Fourier Transformation module to create a series containing fractional frequencies as given below.

$$V_{in} = A \sum_{f_i=f_1}^{f_N} \sin(2\pi f_i t) \qquad (13)$$

A XC-HR50 Sony camera with a capture speed of 120 frames/seconds was used to track the particles with LabVIEW capturing all data via a 1410 NI frame grabber board. Initial attempts to run the experiment resulted in large amplitude excursions of the charged dust grains due to the large amplitudes present within the applied DC oscillation. This made it impossible to resolve small vibration (small amplitude) oscillations of the dust particles. Exchange of particles also sometimes occurred, leaving the bottom particle on top and the top particle on bottom. These issues were eliminated by chopping the waveform into increments of 1 to 3 Hz, providing the resolution necessary to observe small vibrations. Figure 6 shows a FFT of the final data and the resonant frequency for each of the two particles.

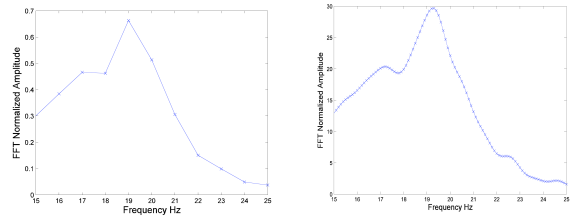

**Figure 6a.** FFT analysis of the experimental data showing a resonant frequency of approximately 19 Hz. **Figure. 6b.** FFT analysis of higher resolution data showing a resonant frequency of approximately 19.2 Hz.

## IV.  RESULTS AND DISCUSSION

Applying a modulated, external DC signal to the lower electrode allowed the resonant frequency of the particle chain system to be determined by employing the techniques given in this paper. The $\beta$ given in Eq. 1 (the Epstein coefficient) should also be able to be determined via the described technique since it is related to the damping coefficient. Once accomplished, the results can be compared with other experimental methods currently in use to determine efficacy of the method [10]. Initial results assuming a diameter for the bottom particle of 8.89 μm produced an apparent diameter for the top particle of 7.80 μm. Since the particles employed in this experiment have a standard deviation of 0.09 μm, this preliminary data will need to be reexamined using finer scanning detail. However, if the technique can be shown as viable, it will provide a simple method for experimentally determining the particle mass and charge as well as measuring the wakefield potential.